\documentclass{emulateapj}
\usepackage{apjfonts}

\def\hi{\ifmmode {\rm H}\,{\sc i}~ \else H\,{\sc i}~\fi}

\slugcomment{Submitted to ApJL}

\shorttitle{Growth of galaxy groups since $z\sim 1$}
\shortauthors{Williams et al.}

\begin{document}

\title{A direct measurement of hierarchical growth in galaxy groups since $z\sim 1$$^\star$}

\author{Rik J. Williams,
        Daniel D. Kelson$^\dagger$,
	John S. Mulchaey,
	Alan Dressler$^\dagger$,
        Patrick J. McCarthy$^\dagger$,
        Stephen A. Shectman}

\affil{Carnegie Observatories, 813 Santa Barbara Street, Pasadena, CA 91101, USA}
\email{williams@obs.carnegiescience.edu}
\altaffiltext{$\dagger$}{Visiting astronomer, Kitt Peak National Observatory, National
Optical Astronomy Observatory, which is operated by the Association of
Universities for Research in Astronomy (AURA) under cooperative
agreement with the National Science Foundation.}
\altaffiltext{$\star$}{This paper includes data gathered with the 6.5 meter Magellan 
Telescopes located at Las Campanas Observatory, Chile.}

\begin{abstract}
We present the first measurement of the evolution of the galaxy group stellar
mass function (GrSMF) to redshift $z\ga 1$ and low masses
($M_\star>10^{12}$\,M$_\odot$). Our results are based on early data from the 
Carnegie-Spitzer-IMACS (CSI) Survey, utilizing low-resolution spectra and 
broadband optical/near-IR photometry to measure redshifts for a $3.6\mu$m
selected sample of 37,000 galaxies over a 5.3\,deg$^2$ area to $z\sim 1.2$.
Employing a standard friends-of-friends algorithm for all
galaxies more massive than $\log M_\star/M_\odot=10.5$, we find a total of $\sim 4000$ groups.
Correcting for spectroscopic incompleteness (including slit collisions), we build
cumulative stellar mass functions for these groups in redshift bins at $z>0.35$,
comparing to the $z=0$  and $z>0$  mass functions from various group and
cluster samples.  Our derived mass functions match up well with $z>0.35$ X-ray
selected clusters, and strong evolution is evident at all masses over the
past 8\,Gyr.  Given the already low level of star formation activity in galaxies
at these masses, we therefore attribute most of the observed growth in the GrSMF 
to group-group and group-galaxy mergers, in accordance with qualitative notions
of hierarchical structure formation.
Given the factor $3-10$ increase in the number density of groups and clusters
with $M_\star>10^{12}$\,M$_\odot$ since $z=1$ and the strong anticorrelation between
star formation activity and environmental density, this late-time growth in group-sized
halos may therefore be an important contributor to the structural and star-formation
evolution of massive galaxies over the past 8\,Gyr.
\end{abstract}


\section{Introduction}
Just as the relative roles of genetics and environment in shaping 
human populations are hotly debated, so are the analogous processes argued 
among those who study the evolution of galaxies.
The ``nature'' of a galaxy is thought to be primarily 
determined by the mass of its dark matter halo, since in isolation the 
collapse of a given amount of matter should give rise to descendents with
a consistent set of properties \citep{berlind02}.
However, like humans, galaxies don't often
mature in isolation: their ``nurturing'' is brought about through gas 
inflows from the extended cosmic web, mergers with other galaxies, and even
direct interactions with hot, dense intergalactic gas \citep[e.g. ram
pressure stripping;][]{koopmann04,crowl05}.  While it
is well-established that galaxies' morphologies \citep{dressler80,wilman09},
colors \citep{cooper07,patel09}, and star-formation properties 
\citep{kauffmann04,patel11,quadri12} differ in
overdense and underdense environments, the reasons behind these correlations 
are still uncertain.

By targeting mass-selected galaxy samples in a range of environments, the role
of ``neighborhood'' can be disentangled from that of mass.  
On the extremes, void galaxies which evolve essentially in isolation 
tend to exhibit stable, gas-rich, star-forming
configurations \citep{grogin00}, while those in rich 
clusters are overwhelmingly quiescent and have morphologies which suggest
frequent high-speed interactions or mergers \citep[e.g.,][]{farouki81}.  Although such
environments provide invaluable laboratories for the study of \emph{ex-} and
\emph{in-situ} galaxy evolution respectively, they represent rare and extreme
neighborhoods.  Most of the galaxy population in fact resides in the proverbial
suburbs: relatively small galaxy groups containing anywhere between two to a
few dozen significant members \citep{eke04}. Groups' influence
on their member galaxies may be dramatic, like quenching star formation 
by keeping cold gas from reaching galaxies
\citep[e.g.,][]{vdb08} and/or triggering bursts of star formation through
slower galaxy-galaxy interactions than clusters
\citep{larson78}.  In turn, the assembly of groups (and the quenching
processes therein) since $z=1$ could be a major factor in the decline of the
cosmic star formation rate over the same time period \citep{lilly96,madau96}.

Despite the ubiquity of groups, homogeneous samples are notoriously difficult
to assemble beyond the nearby universe ($z\gtrsim 0.4$).  Since each
group contains only a few galaxies spread over $\sim 1$ Mpc or more, standard
cluster selection techniques (e.g. angular overdensities exhibiting a
well-defined red sequence) cannot be applied.  Comprehensive spectroscopic
surveys over wide areas are the only reliable way to weed out groups from
interlopers; however, at high redshifts optically-selected spectroscopic
surveys are biased toward UV-bright, highly star-forming galaxies at the
expense of the massive red galaxies which dominate dense environments.
An \emph{infrared-selected} survey, effectively producing a
mass-limited sample at $z\sim 1$, is therefore necessary to produce an unbiased
census of massive galaxies at these redshifts \citep{vdokkum06,kelson12}.  

The Carnegie-Spitzer-IMACS (CSI) Survey\footnote{More information at \url{http://csi.obs.carnegiescience.edu}}, currently underway at the
Magellan-Baade 6.5m telescope in Chile, has been specifically designed to
characterize massive galaxies and their environments up to $z\sim 1.4$.  Here we present a
measurement of the evolution of the \emph{group} stellar mass function (GrSMF),
to lower group masses and higher redshifts than has been previously achieved, using
early data from CSI.  Cosmological parameters $h=0.7$, $\Omega_m=0.3$, and
$\Omega_\Lambda=0.7$ are assumed throughout.

\section{Observations and spectral fitting} \label{sec_data}
In a companion paper \citep{kelson12}, we describe the CSI observing setup and
strategy, data reduction, and SED fitting; a brief summary follows.  We
employ a simple $3.6\mu$m flux limit of $m_{\rm AB}<21$ to select galaxies from
three of the Spitzer Wide-Area Extragalactic Survey (SWIRE) fields: ELAIS-S1,
\emph{Chandra} Deep Field South, and XMM-Large Scale Structure (XMM-LSS),
excluding stars with simple optical/IR color cuts. The $3.6\mu$m band
lies in the rest-frame near-IR for galaxies at $z=0.5-1.5$, resulting
in an approximately mass-selected sample unbiased by dust and star
formation.  The first two years of observations 
focused primarily on the XMM-LSS field, giving it the 
best spectroscopic and photometric coverage at present; 
here we focus only on data from this 5.3\,deg$^2$ field in our analysis.

Targets were observed in multi-slit, nod-and-shuffle mode \citep{glazebrook01}
with a multiplexing efficiency of $\sim 1800$ objects per $28^\prime$ diameter
mask, using the low-dispersion prism (LDP) on the Inamori-Magellan Areal Camera
and Spectrograph \citep[IMACS;][]{dressler11} in 2009. In 2010, a second pass
was mostly completed with the new uniform-dispersion prism (UDP) which provides
superior resolution beyond $\lambda>7500$\AA. Some overlap between the LDP- and
UDP-observed samples was included for cross-calibration and testing; overall,
two passes have provided $\sim 40$\% completeness.  The IMACS spectra are 
supplemented with moderate-depth $J$ and $K$
imaging from the NOAO Extremely Wide-Field InfraRed Mosaic (NEWFIRM)
and optical photometry from the public CFHT Legacy Survey.

The spectra were optimally extracted and combined, and prism and
broadband IR data were jointly fit using a generalized set of starburst
models described by \citet{kelson12}. With a series of benchmarks --- 
spectroscopic redshifts in this field \citep[VVDS;][]{lefevre05}, galaxies
observed with both the LDP and UDP, 
and comparing prism redshifts of physically associated galaxy pairs \citep[cf.][]{quadri10}
--- we infer
typical redshift uncertainties of $\Delta z/(1+z)\sim 0.005-0.015$ at $z<1$.
Notably, unlike photometric redshifts, these errors are comparable for
red and blue galaxies since the prism spectrum pinpoints both the 4000\AA\ break
and emission lines.
After excluding objects with poor SED fits (typically due to strong AGN
components, bad photometry, and/or incorrect slit placement) the final 
catalog contains about 37,000 galaxy redshifts over 5.3 deg$^2$, 
with an effective stellar mass limit of about $M_\star>3\times 10^{10}$M$_\odot$ 
at $z\sim 1$.

\section{Galaxy groups at $0.5<z<1.2$}
\subsection{Finding groups with friends-of-friends}
Groups typically appear as associations of two to a few tens of galaxies over
an area of a few hundred projected kpc; accurate redshifts and large areas are
therefore critical to building  robust group samples. We employ a simple
friends-of-friends \citep[FoF;][]{huchra82} algorithm to find groups in the CSI catalog, limiting
the sample to galaxies with $M_\star>3.2\times 10^{10}$\,M$_\odot$, the CSI
mass limit for red galaxies at $z\sim 1.2$. Even at the low end our redshift
uncertainty ($\sim 0.5$\%) is larger than the expected velocity dispersion
of most virialized groups and clusters. Redshift uncertainties were estimated
(using the techniques mentioned in Section~\ref{sec_data})
as a function of $r$-band magnitude and the quadrature sum of each galaxy
pair's $1\sigma$ uncertainties was adopted
as the redshift linking length.  We used a transverse linking length of
$1.0/(1+z)$\,Mpc (matching  the \citealt{yang07} length of 1\,Mpc for SDSS groups). The
mean group redshift was also adjusted as new members were added, and previous
group members falling outside 1 $\sigma_z$ from the new mean were removed from
the group.  In cases where the members of two groups overlapped, the two were
merged into a single group, the linking length recalculated, and the membership
adjusted accordingly.

\begin{figure*}
\plottwo{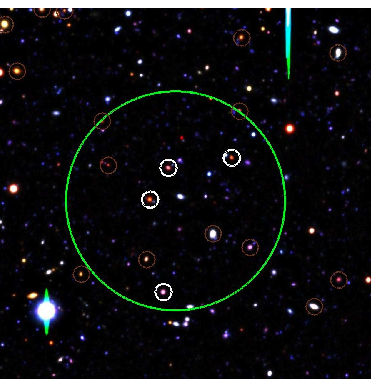}{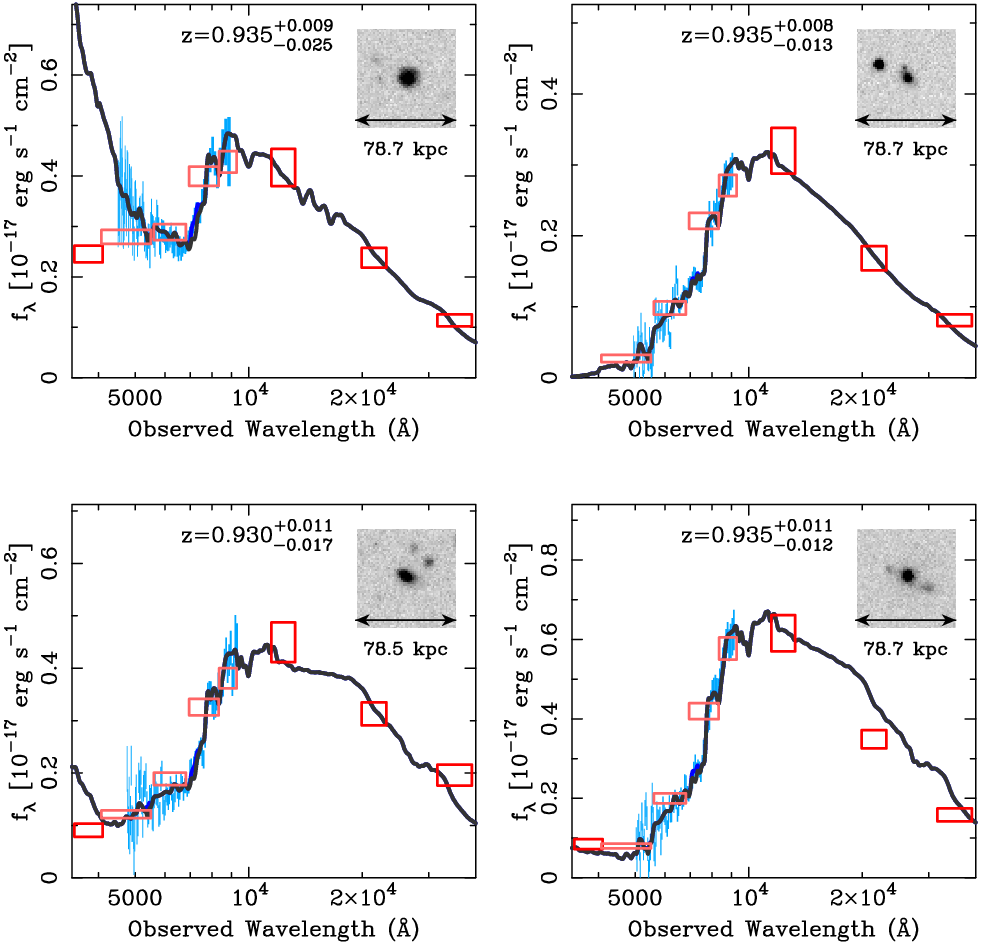}
\caption{\emph{Left panel:} Example of a group at $z=0.935$ with confirmed members
circled in white, shown in a $3^\prime \times 3^\prime$ (1.4 Mpc on a side)
region. Background and foreground galaxies in our spectroscopic sample are circled in brown; the
large green circle is $\pi/2$ times the group's estimated mean harmonic radius.
\emph{Right panels:} Spectral energy distributions of the four confirmed group
members. The cyan histogram shows the prism data, red boxes denote broadband
photometry and uncertainties, and the black line is the best-fit model
for each galaxy. All of these objects show strong 4000\AA\ breaks, indicating
their stellar populations are largely evolved, though one exhibits
current star formation. 
\label{fig_grpex} }
\end{figure*}

Ultimately this iterative FoF process found 1551 groups in the CSI
catalog containing a total of 4140 confirmed galaxies above the mass limit, or
an estimated 13000 galaxies after correcting for incompleteness (as described
in the next section). This correction is somewhat larger than that implied by 
the mean completeness of $\sim 30-40$\% \citep{kelson12}, because
groups and clusters are relatively dense environments that are difficult to
sample with multislit spectroscopy.  Figure~\ref{fig_grpex} shows a $VzJ$
color image of one representative group at $z=0.935$ along with spectral energy
distributions (SEDs) of its four confirmed members. These galaxies are circled
in white in the image, and with this particular color scale appear to have
consistent colors. Three of the member galaxy SEDs look fairly similar,
with primarily evolved stellar populations, while the fourth 
is blue and star forming.  This illustrates the high fraction of 
massive evolved galaxies even in relatively poor groups at high
redshifts.

\subsection{Incompleteness and projection effects}
Even with the redshift accuracy of CSI,
two effects will introduce serious biases if not taken into account: spurious
groups introduced by projected alignments of galaxies, and group members not
included in the sample due to the minimum slit spacing of IMACS and
yet-incomplete coverage. We correct for the latter effect by creating
mock galaxy catalogs based on the target source density and
flux distribution (i.e., that of all objects above the IRAC flux limit),
and estimating the fraction of sources on which slits could be placed $f(\rho)$ as
a function of projected density. For a given region of the image,
the underlying source number density is then approximated by $n_{\rm true}=n_{\rm obs}/f(\rho)$.

To estimate the contribution of spurious groups due to chance projections,
we assigned random redshifts to the galaxies in the sample
(keeping the same distribution of source densities)
and re-ran the group finding algorithm.  Not
surprisingly, the likelihood of chance projections decreases with increasing
group mass since a higher concentration of galaxies in a given region is more
likely to be a ``true'' group or cluster; this contamination fraction varies from 60\% at
the low end ($\log M_\star/M_\odot\sim 11.5$) to 10\% for groups an order of
magnitude more massive; over the full group sample the mean contaminant fraction is about 
40\%.  We therefore only consider group stellar masses above $\log (M_\star/M_\odot)>11.7$, 
where the contribution from spurious groups is below 50\%.

\subsection{Calculating the cumulative group stellar mass functions}
The CSI group catalog was conservatively selected from a galaxy sample which is
complete to $\log (M_\star/M_\odot)>10.5$ at $z< 1.2$ for both red and blue
galaxies. Calculation of the GrSMFs is therefore straightforward,
simply a matter of summing the number of groups detected in mass and redshift
bins and dividing by the effective survey volume (taking into account area not
covered due to gaps between masks, bright stars, etc.).  Above the adopted mass limit of 
groups above $\log (M_\star/M_\odot)=11.7$, there are 686 groups.  
We subdivided the sample into redshift
bins $z=0.35-0.55$, $0.55-0.9$, and $0.9-1.2$.  Our cumulative GrSMFs are binned such that
the first (highest-mass) bin contains three galaxies, and the number
of groups in each successively lower-mass bin is incremented by either one or a multiplicative
factor of 1.1, whichever is greater; this
provides a roughly constant binning in $\log N$ at low masses.  Finally, to correct for
projection effects the GrSMF of the previously-described ``randomized''
(i.e. spurious) groups was calculated the same way and subtracted.  

To check the concordance of these GrSMFs with those found in other
studies at lower redshifts and higher masses, we adopted two complementary samples
from the literature: the Sloan Digital Sky Survey (SDSS) group catalog by
\citet{yang07}, and massive X-ray selected clusters at $z\sim 0$ and $z>0.35$
from the ROSAT All-Sky and 400 Degree (400d) surveys
\citep{burenin07,vikhlinin09}. 400d includes 36 clusters at $z>0.35$ and
49 at $z\la 0.2$ detected in 400 deg$^2$ 
of \emph{ROSAT} PSPC observations and followed up with \emph{Chandra} to obtain
total virial masses.  To create a $z\sim 0$ sample comparable to CSI,
we chose a highly-restricted subset of groups from the \citet{yang07} ``Sample II'' catalog
with at least two galaxies of $\log (M_\star/M_\odot)>10.5$ between
$0.025<z<0.075$ (the upper bound being the redshift where SDSS is complete
at these masses), and calculated the cumulative group mass function. In contrast,
the 400d survey provides an X-ray selected cluster sample given in virial, not
stellar, masses. We therefore recomputed their mass functions in terms
of $M_{500}$ using the volumes in Figure 11 of \citet{vikhlinin09} and 
transformed their virial masses to stellar masses with the \citet{giodini09} relations. 

If the group masses have substantial uncertainties, the observed mass function will be
biased toward larger masses (due to the steepness of the GrSMF). This effect is
small for SDSS and 400d, but significant for CSI which has relatively large mass errors. To correct
for this, we convolved the SDSS mass function with CSI's
expected stellar mass error distribution, and thereby estimated the shift
in the CSI GrSMF as a function of group stellar mass. All CSI GrSMFs were
corrected accordingly, with a mass shift of $\sim 0.5$\,dex.

Figure~\ref{fig_mf} shows the CSI, SDSS, and \citet{vikhlinin09} mass functions
overplotted, all in terms of $\log M_\star/M_\odot$. As an additional check, we include the
HIFLUGCS X-ray selected cluster mass function from \citet{reiprich02},
transforming their $M_{200}$ values to $M_\star$ with relations given by
\citet{lagana11}  (which were specifically computed from HIFLUGCS). Poisson
errors are shown on the CSI data points; cosmic variance is not included, but
using the \citet{trenti08} Cosmic Variance Calculator we estimate it will add an 
additional 10-20\% (greatest in the $0.35<z<0.55$ bin and at
the highest masses) to the number density uncertainty; however, for the most 
part the Poisson uncertainties dominate. 
Despite the major selection differences, the SMFs
of the $z\sim 0$ optical and X-ray samples are in good agreement, suggesting
that the adopted transformations from virial to stellar masses are reasonably
robust.

\section{Discussion}
\subsection{Concordance with massive X-ray selected clusters}
The effective upper mass limit of our GrSMFs where we ``run out'' of volume are,
in the realm of galaxy clusters, not particularly massive: at $M_\star \sim 10^{12.5}$\,M$_\odot$ 
(above which we only have 12 groups in our sample), in the literature they would 
typically be called ``rich groups'' or ``poor clusters.''  The lack of rich CSI clusters is simply
a result of the limited area covered by CSI: while 5.3\,deg$^2$ (and the
ultimate goal of 15\,deg$^2$) is very large for a $z\sim 1$ galaxy survey, the
richest, most massive clusters are rare enough that much larger areas are
needed to find significant numbers of them at low to intermediate redshifts.
As noted in the previous section, such samples are provided by large X-ray
surveys covering hundreds of square degrees; here we have adopted the 400d
survey \citep{vikhlinin09} as a principal comparison sample.  

\begin{figure}
\plotone{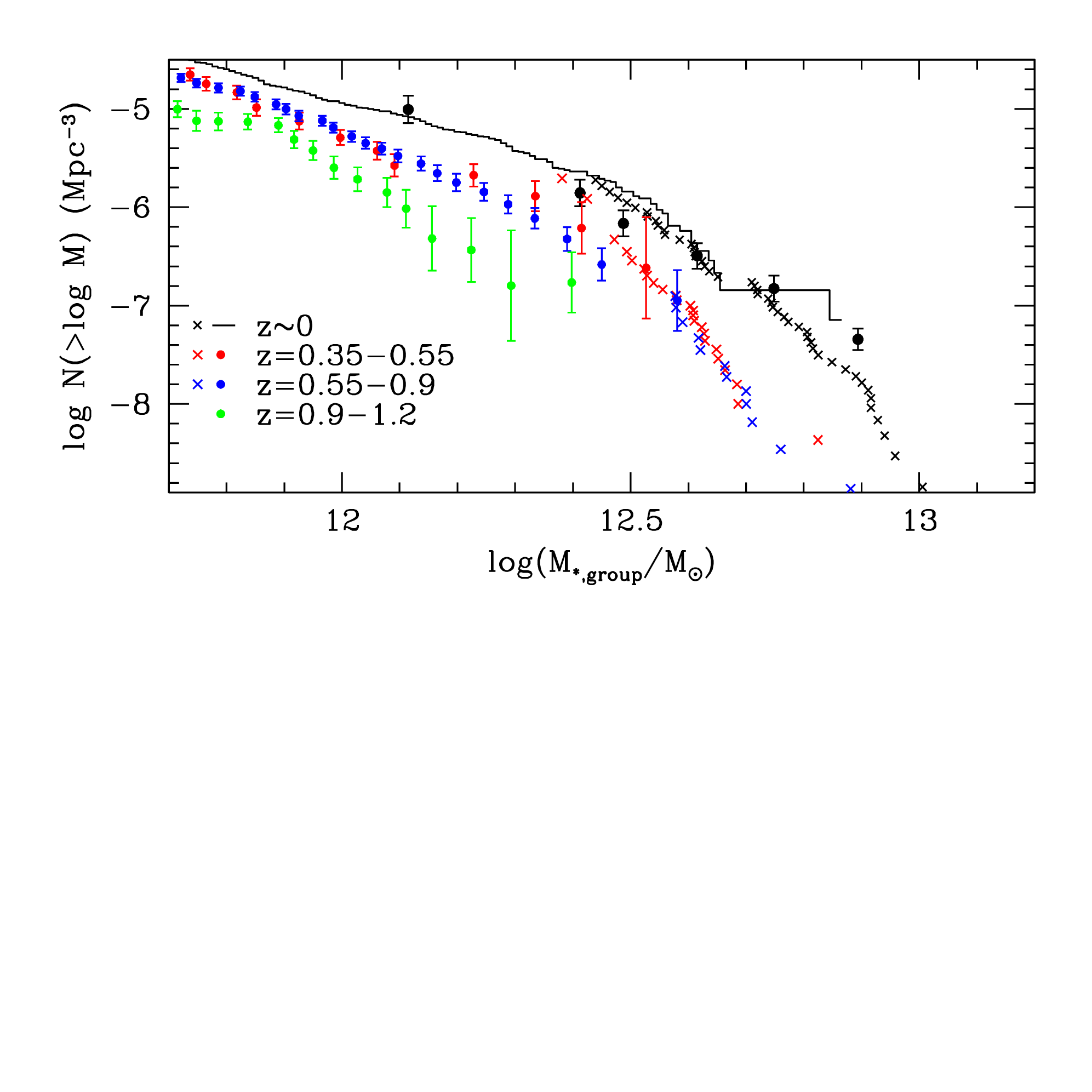}
\caption{Evolution of the cumulative GrSMF from $z=0-1$. Data from CSI at
low-, medium-, and high-redshifts are shown as red, blue, and green
circles (with Poisson error bars) respectively. The black histogram shows the
$z\sim 0$ mass function derived from the \citet{yang07} SDSS group catalog,
while the black circles are X-ray selected $z\sim 0$ clusters from HIFLUGCS
\citep{reiprich02}. Crosses show mass functions from the \citet{vikhlinin09}
400d survey converted from virial to stellar masses, with colors corresponding
to the same redshift bins as the CSI and SDSS samples. All $z\sim 0$ estimates of
the mass function match up well despite the different selection techniques, and
strong evolution in both group and cluster number densities is evident across
the plotted redshift range.
\label{fig_mf} }
\end{figure}

Figure~\ref{fig_mf} shows that the cluster mass functions of \citet{vikhlinin09}
(transformed to stellar masses by the \citealt{giodini09} conversion factors)
not only pick up more or less where CSI leaves off, but also represent a smooth
continuation of the CSI GrSMFs to a factor $\sim 2$ higher mass.  This
is the first demonstration of the connection between group and
cluster SMFs at these high redshifts.  While
not a particularly surprising result, it provides further evidence that
our observing strategy and group selection methods are robust, and that the
groups found by CSI bridge a key gap between individual galaxies and
massive clusters in the distribution of dark matter halos. 
Although deep X-ray and IR observations provide compelling evidence of 
massive clusters at $z>1$ \citep{mccarthy07,papovich10,rosati09}, no 
\emph{homogeneously-selected} sample currently exists at these redshifts;
indeed, the rapid decline in cluster abundances at high stellar masses
seen in Figure~\ref{fig_mf}
(if it is similarly steep at $z=0.9-1.2$) suggests that such objects
are extremely rare. The most common progenitors of low-$z$ clusters
must therefore lie at group masses at $z>1$.

\subsection{Hierarchical growth over the past 8 Gyr}
The combined CSI and 400d mass functions show strong evolution over the
redshift range $z=0.35-1.2$, with much of the evolution occurring between the
$z\sim 1$ and $z\sim 0.6$ redshift bins.  Similarly, there's another
significant increase in number density between the lowest-redshift bin of CSI and
the $z=0$ groups and clusters.  In the two lowest-redshift CSI bins there appears to
be only marginal evolution between the mass functions; this may be a result of 
cosmic variance or residual mass uncertainties. Overall, however, the observed evolution appears to
qualitatively reflect hierarchical structure formation: the ranks of massive groups and
clusters ($\log M_\star/M_\odot\ga 12.2$) grow strongly over this
redshift interval, while the number density of lower-mass groups is more
constant (presumably because new groups are formed from below our mass limit as
others merge into the more massive clusters).  Although these hierarchical trends
can be seen in the 400d cluster data alone, CSI and SDSS demonstrate that this
growth continues to masses $\sim 0.5$\,dex lower. 

\begin{figure}
\plotone{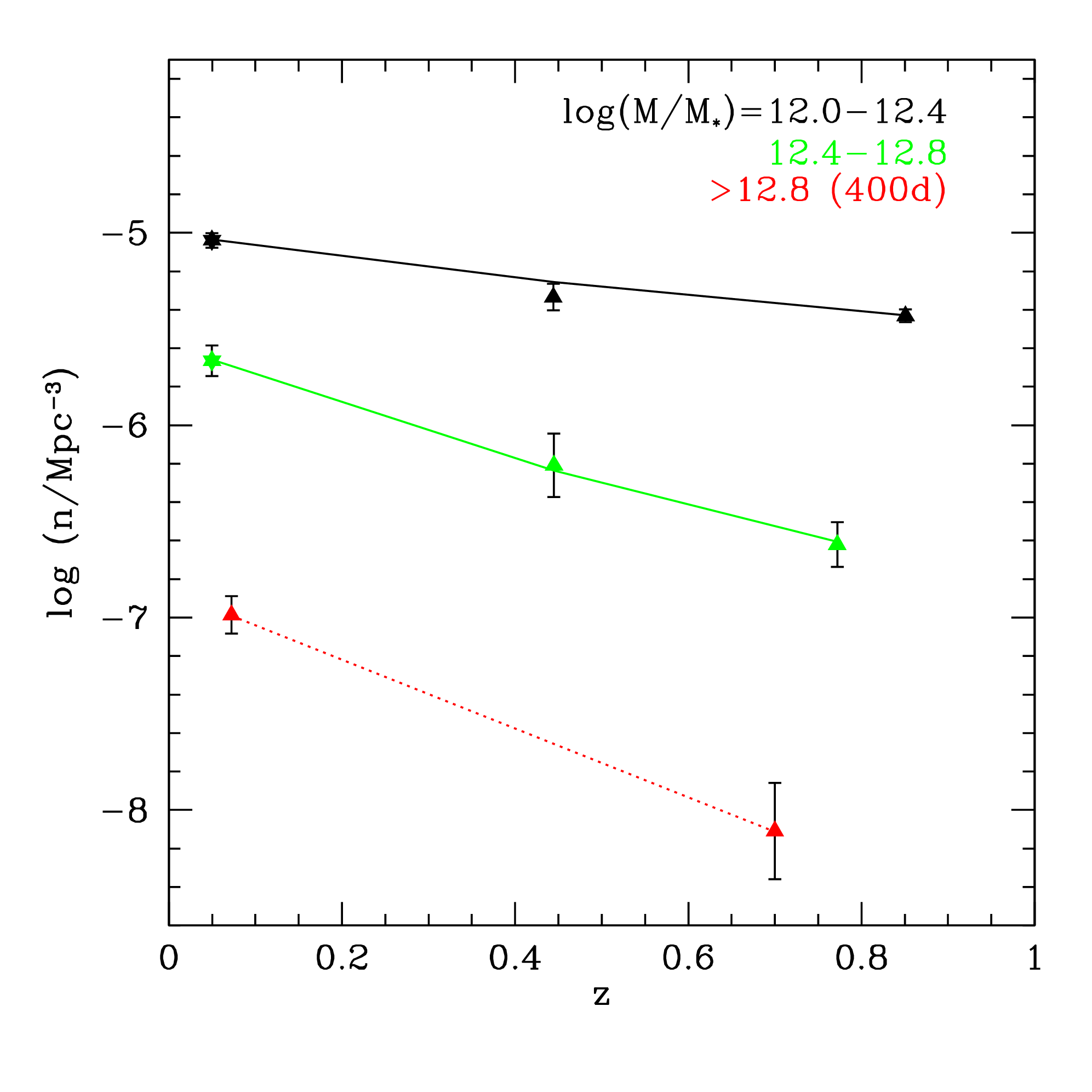}
\caption{
Number density of groups in three stellar mass bins. For the two lower
mass bins, data at $z\sim 0$ and $z>0.25$ are taken from SDSS and
CSI respectively, while the most massive clusters are from the low- and high-$z$ samples of
\citet{vikhlinin09}, converted to stellar masses. Lines are power-law fits of the form
$n\sim (1+z)^\alpha$, with $\alpha=-1.6\pm 0.2$, $-4.2\pm 0.7$, and $-5.6\pm 1.4$
in order of increasing mass; due to potential systematic offsets between the optically-
and X-ray-selected samples, the latter is less reliable and we show it as a dotted line. The hierarchical 
buildup of groups, where those with higher masses grow more rapidly,
is evident in this figure.
\label{fig_mbins} }
\end{figure}

Due to the small number of high-mass clusters, the mass functions in 
Figure~\ref{fig_mf} are shown as cumulative number densities. To  
better illustrate the observed hierarchical growth, another projection of the GrSMF
evolution is shown in Figure~\ref{fig_mbins}. Here the number
densities of galaxies in three mass bins (two from CSI/SDSS and one from 
400d) are shown as a function of redshift.  The difference between
low- and high-mass groups is striking: at $\log (M_\star/M_\odot)=12-12.4$
the abundance of groups is relatively flat, but at masses $\sim 0.4$\,dex higher
the number density declines far more rapidly with increasing redshift.  More quantiatively,
we fit a power-law $n\sim (1+z)^\alpha$ and find $\alpha=-1.6\pm 0.2$
for groups with $12.0<\log M<12.4$ and $\alpha=-4.2\pm 0.7$ at $12.4<\log M<12.8$; 
the difference in slopes is therefore significant at the $3.5\sigma$
level.  X-ray clusters may exhibit marginally faster growth ($\alpha=-5.6\pm 1.4$),
but due to the large uncertainties and potential systematics in their estimated stellar masses 
we cannot determine whether this steepening trend continues to $\log M_\star> 12.8$.

\section{Summary and future work}
By employing low-resolution prism spectroscopy to obtain accurate redshifts for
a mass-complete galaxy sample over 5.3\,deg$^2$, we have assembled the most
comprehensive catalog of 686 galaxy groups down to a stellar mass limit of
$10^{11.7}$\,M$_\odot$ and up to redshifts $z\sim 1.2$.  The CSI GrSMFs
follow a smooth continuation of the X-ray selected cluster mass
functions of \citet{vikhlinin09} when these are transformed to stellar masses.
Most notably, the number density of high-mass groups and clusters increases
more rapidly than low-mass groups; since the galaxies in these objects
exhibit little ongoing star formation, hierarchical buildup through
mergers and accretion must dominate their growth.

This result illustrates the power of large prism surveys like CSI to
effectively bridge the gap between galaxy- and cluster-mass halos, 
allowing studies of an ubiquitous galaxy environment. As noted before,
this result is based on an early sample comprising about one-third of the total
CSI area and 25\% of the expected spectra. While these early data provide a
qualitative picture of hierarchical growth in group stellar mass, forthcoming
CSI data which attain improved completeness over the full 15\,deg$^2$ area will
allow us to robustly track the stellar-mass growth of groups and clusters, and for the first 
time directly test theoretical models in the gap between galaxy and cluster-mass
dark matter halos. Additionally,
deep X-ray observations (and/or stacks of existing shallower data) in the CSI fields
will allow a direct comparison of stellar and virial masses for X-ray detected
groups, providing an unprecedented window into the underlying halo growth.

\acknowledgments
We thank Ryan Quadri for helpful discussions and comments on the manuscript.
In addition, R.J.W.~gratefully acknowledges support from the Carnegie Institution for
Science and NSF grant AST-0707417.


\clearpage

\begin{thebibliography}{}
\bibitem[Berlind \& Weinberg(2002)]{berlind02} Berlind, A. \& Weinberg, D.~H.
2002, \apj, 575, 587
\bibitem[Burenin et al.(2007)]{burenin07} Burenin, R.~A., Vikhlinin, A.,
Hornstrup, A., et al.~2007, \apjs, 172, 561
\bibitem[Cooper et al.(2007)]{cooper07} Cooper, M.~C., Newman, J.~A., 
Weiner, B.~J., et al.~2007, \mnras, 376, 1445
\bibitem[Crowl et al.(2005)]{crowl05} Crowl, H.~H., Kenney, J.~D.~P.,
van Gorkom, J.~H., \& Vollmer, B.~2005, \aj, 130, 65
\bibitem[Dressler(1980)]{dressler80} Dressler, A.~1980, \apj, 236, 351
\bibitem[Dressler et al.(2011)]{dressler11} Dressler, A., Bigelow, B., Hare, T.,
et al.~2011, \pasp, 123, 288
\bibitem[Eke et al.(2004)]{eke04} Eke, V.~R., Baugh, C.~M., Cole, S., et 
al.~2004, \mnras, 348, 866
\bibitem[Farouki \& Shapiro(1981)]{farouki81} Farouki, R.~T. \& Shapiro, 
S.~L.~1981, \apj, 243, 32
\bibitem[Giodini et al.(2009)]{giodini09} Giodini, S., Pierini, D., 
Finoguenov, A., et al.~2009, \apj, 703, 982
\bibitem[Glazebrook \& Bland-Hawthorn(2001)]{glazebrook01} Glazebrook, K.
\& Bland-Hawthorn, J.~2001, \pasp, 113, 197
\bibitem[Grogin \& Geller(2000)]{grogin00} Grogin, N.~A. \& Geller, M.~J.~2000,
\aj, 118, 2561
\bibitem[Huchra \& Geller(1982)]{huchra82} Huchra, J.~P. \& Geller, M.~J.~1982,
\apj, 257, 423
\bibitem[Kauffmann et al.(2004)]{kauffmann04} Kauffmann, G., White, S.~D.~M.,
Heckman, T.~M., et al.~2004, \mnras, 353, 713
\bibitem[Kelson et al.(2012)]{kelson12} Kelson, D.~D., et al.~2012, \apj, submitted
\bibitem[Koopmann \& Kenney(2004)]{koopmann04} Koopmann, R.~A. \& Kenney, 
J.~D.~P.~2004, \apj, 613, 866
\bibitem[Lagan\'a et al.(2011)]{lagana11} Lagan\'a, T., Zhang, Y.-Y., Reiprich,
T.~H., \& Schneider, P.~2011, \apj, in press (arXiv:1108.3678)
\bibitem[Larson \& Tinsley(1978)]{larson78} Larson, R.~B. \& Tinsley, B.~M.~1978,
\apj, 219, 46
\bibitem[Le Fevre et al.(2005)]{lefevre05} Le Fevre, O., et al.~2005, \aap, 439, 845
\bibitem[Lilly et al.(1996)]{lilly96} Lilly, S.~J., Le Fevre, O., Hammer, F.,
\& Crampton, D.~1996, \apj, 460, L1
\bibitem[Madau et al.(1996)]{madau96} Madau, P., Ferguson, H.~C., Dickinson,
M.~E., et al.~1996, \mnras, 283, 1388
\bibitem[McCarthy et al.(2007)]{mccarthy07} McCarthy, P.~J., Yan, H., Abraham,
R.~G., et al.~2007, \apj, 664, L17
\bibitem[Papovich et al.(2010)]{papovich10} Papovich, C., Momcheva, I., Willmer,
C.~N.~A., et al.~2010, \apj, 716, 1503
\bibitem[Patel et al.(2009)]{patel09} Patel, S.~G., Kelson, D.~D., Holden, B.~P.,
et al.~2009, \apj, 694, 1349
\bibitem[Patel et al.(2011)]{patel11} Patel, S.~G., Kelson, D.~D., Holden, B.~P.,
Franx, M., \& Illingworth, G.~D.~2011, \apj, 735, 53
\bibitem[Rosati et al.(2009)]{rosati09} Rosati, P., Tozzi, P., Gobat, R., et
al.~2009, \aap, 508, 583
\bibitem[Quadri \& Williams(2010)]{quadri10} Quadri, R.~F., \& Williams, 
R.~J.~2010, \apj, 725, 794
\bibitem[Quadri et al.(2012)]{quadri12} Quadri, R.~F., Williams, R.~J., Franx, M.,
\& Hildebrandt, H.~2012, \apj, 744, 88
\bibitem[Reiprich \& B\"ohringer(2002)]{reiprich02} Reiprich, T. \& B\"ohringer,
H.~2002, \apj, 567, 716
\bibitem[Trenti \& Stiavelli(2008)]{trenti08} Trenti, M., \& Stiavelli, M.~2008,
\apj, 676, 767
\bibitem[van den Bosch et al.(2008)]{vdb08} van den Bosch, F.~C., Aquino, D.,
Yang, X., et al.~2008, \mnras, 387, 79
\bibitem[van Dokkum et al.(2006)]{vdokkum06} van Dokkum, P.~G., Quadri, R.,
Marchesini, D., et al.~2006, \apj, 638, L59
\bibitem[Vikhlinin et al.(2009a)]{vikhlinin09} Vikhlinin, A., Burenin, R.~A., 
Ebeling, H., et al.~2009, \apj, 692, 1033
\bibitem[Wilman et al.(2009)]{wilman09} Wilman, D.~J., Oemler, A., Jr., 
Mulchaey, J.~S., et al.~2009, \apj, 692, 298
\bibitem[Yang et al.(2007)]{yang07} Yang, X., Mo, H.~J., van den Bosch,
F.~C., et al.~2007, \apj, 671, 153
\end{thebibliography}
\end{document}